\documentclass[a4paper,12pt]{article}

\usepackage{hyperref, amsmath, amssymb}
\newif\ifpdf
\ifx\pdfoutput\undefined
\pdffalse 
\else
\pdfoutput=1 
\pdftrue
\fi

\ifpdf
\usepackage[pdftex]{graphicx}
\else
\usepackage{graphicx}
\fi

\usepackage{subfigure}

\ifpdf
\DeclareGraphicsExtensions{.pdf, .jpg, .tif}
\else
\DeclareGraphicsExtensions{.eps, .jpg}
\fi

\begin{document}
\begin{center}
\textbf{Formation of quasi-solitons in transverse confined ferromagnetic film media}

\vspace{1.0cm}
A.A. Serga \footnote{Email address: serha@rhrk.uni-kl.de}

\textit{Technische Universit{\"a}t Kaiserslautern, Department of Physics and Forschungsschwerpunkt MINAS, D - 67663 Kaiserslautern, Germany}

\vspace{0.5cm}
M. Kostylev \footnote{Email address: kostylev@cyllene.uwa.edu.au}

\textit{School of Physics, The University of Western Australia, 35 Stirling Highway, Crawley WA 6009, Australia
\\
and
\\
St.Petersburg Electrotechnical University, 197376, St.Petersburg, Russia}

\vspace{0.5cm}
B. Hillebrands

\textit{Technische Universit{\"a}t Kaiserslautern, Department of Physics and Forschungsschwerpunkt MINAS, D - 67663 Kaiserslautern, Germany}
\end{center}

\vspace{1cm}

\begin{small}
\textbf{Abstract} The formation of quasi-2D spin-wave waveforms in longitudinally magnetized
stripes of ferrimagnetic film was observed by using time- and
space-resolved Brillouin light scattering technique. In the linear regime
it was found that the confinement decreases the amplitude of dynamic
magnetization near the lateral stripe edges. Thus, the so-called effective
dipolar pinning of dynamic magnetization takes place at the edges.

In the
nonlinear regime a new stable spin wave packet propagating along a waveguide
structure, for which both transversal instability and interaction with the side walls of the waveguide
are important was observed. The experiments and a numerical simulation of
the pulse evolution show that the shape of the formed waveforms and
their behavior are strongly influenced by the confinement.
\end{small}

\vspace{1cm}

We report on the observation of a new type of a stable, two-dimensional nonlinear spin wave
packet propagating in a magnetic waveguide structure and suggest a theoretical description of our
experimental findings. Stable two-dimensional spin wave packets, so-called spin wave bullets,
were previously observed, however solely in long and wide samples of a thin ferrimagnetic film
of yttrium-iron-garnet (YIG) \cite{Linear_nonlinear_diffraction_dipolar_spin_waves_yttrium_iron_garnet_films_observed_space-_time-resolved_Brillouin_light_scattering,
Self-Generation_Two-Dimensional_Spin-Wave_Bullets,
Parametric_Generation_Forward_Phase-Conjugated_Spin-Wave_Bullets_Magnetic_Films},
that were practically unbounded in both in-plane directions
compared to the lateral size of the spin wave packets and the wavelength of the carrier spin wave.
In a waveguide structure, where the transverse dimension is comparable to the wavelength, up to
day only quasi one-dimensional nonlinear spin wave objects were observed, which are spin wave
envelope solitons. Here a typical system is a narrow 
({$\simeq$ 1-2}{mm}) stripe of a YIG ferrite film \cite{Collision_properties_quasi-one-dimensional_spin_wave_solitons_two-dimensional_spin_wave_bullets,
Backward-volume-wave_microwave-envelope_solitons_in_yttrium_iron_garnet_films}.
Both for solitons and bullets the spreading in dispersion is compensated by the longitudinal nonlinear
compression. Concerning the transverse dimension, solitons have a cosine-like amplitude
distribution due to the lateral confinement in the waveguide, whereas bullets show a transverse
nonlinear instability compensating pulse widening due to diffraction and leading to transverse
confinement.

Here we report on the observation of a new stable spin wave packet propagating along a waveguide
structure, for which both transversal instability and interaction with the side walls of the waveguide
are important.

The experiments were carried out using a longitudinally magnetized long YIG film stripe of 2.5mm
width and 7$\mu$m thickness. The magnetizing field was 1831Oe. The spin waves were excited by
a microwave magnetic field created with a microstrip antenna of 25$\mu$m width placed across the
stripe and driven by electromagnetic pulses of 20ns duration at a carrier frequency of 7.125GHz.
As is well known the backward volume magnetostatic spin wave (BVMSW) \cite{Overview_electromagnetic_and_spin_angular_momentum_mechanical_waves_ferrite_media}  excited in the
given experimental configuration is able to form both envelope solitons and bullets \cite{Collision_properties_quasi-one-dimensional_spin_wave_solitons_two-dimensional_spin_wave_bullets}, depending
on the geometry. The spatio-temporal behavior of the traveling BVMSW packets was investigated
by means of space- and time-resolved Brillouin light scattering spectroscopy \cite{Physics_Reports}.

\begin{figure}[tpb]
\begin{center}
	\includegraphics[width=13cm]{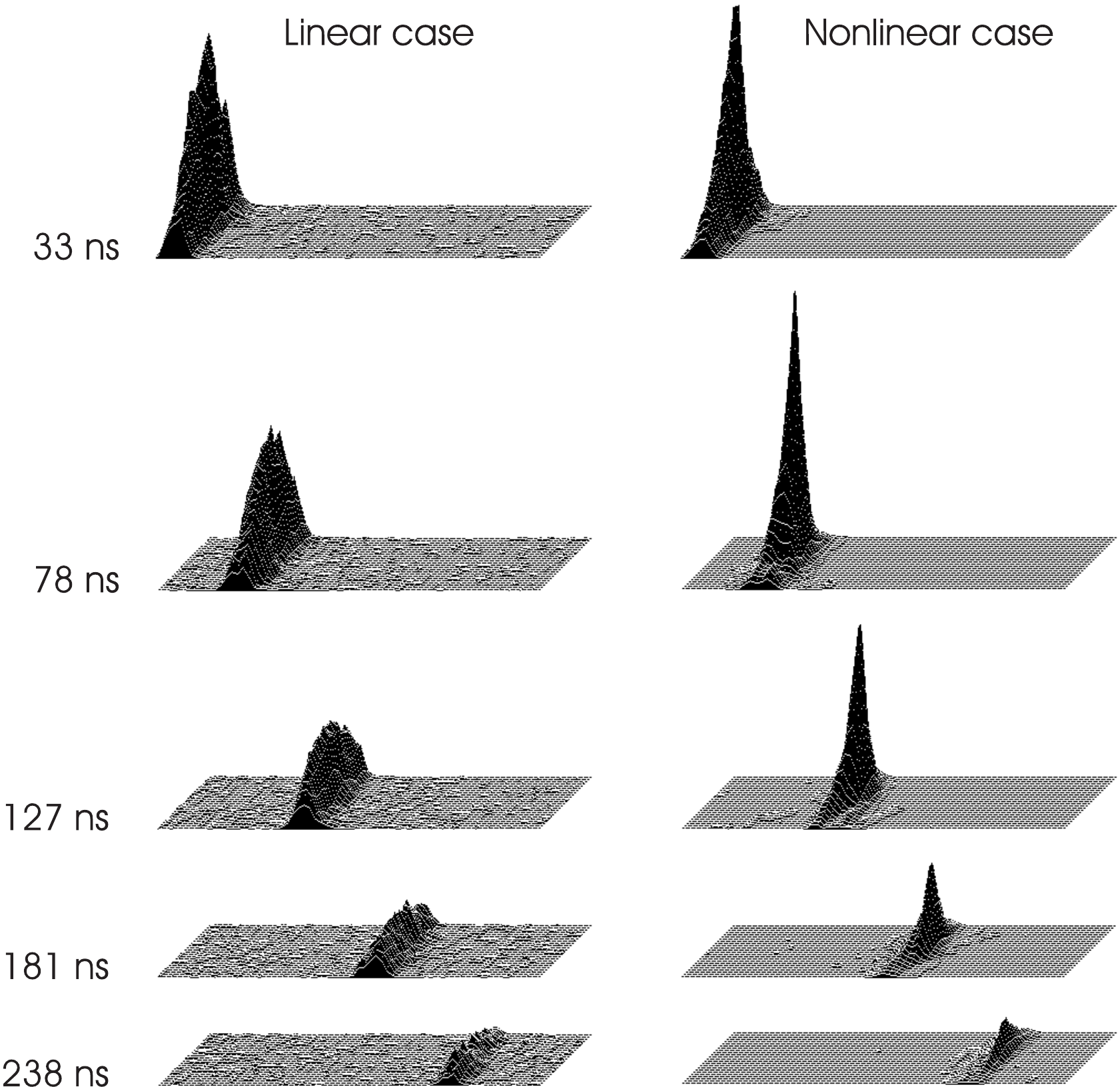} 

	\caption{Bullet formation in the transversally confined yttrium-iron-garnet film.}
		\label{PulseProfiles}
	\end{center}
\end{figure}

The obtained results are demonstrated in Fig.~\ref{PulseProfiles} where the spatial distributions of the intensity of
the spin wave packets are shown for given moments of time. The spin wave packets propagate
here from left to right and decay in the course of their propagation along the waveguide because
of magnetic loss. The left set of diagrams corresponds to the linear case. The power of the driving
electromagnetic wave is 20mW. The right set of diagrams corresponding to the nonlinear case was
collected for a driving power of 376mW.

Differences between these two cases are clearly observed. First of all the linear spin wave packet
is characterized by a cosine-like lateral profile while the cross section of the nonlinear pulse is
sharply modified relative to the linear case and has a pronounced bell-like shape. Second, the
intensity of the linear packet decays monotonically with time while the intensity of the nonlinear
packet initially increases because of its strong transversal compression (see the second diagram
from the top in Fig.~\ref{PulseProfiles}).

Both of these nonlinear features provide clear evidence for the development of a transversal instability
and bullet formation. It is interesting that the bell-like cross-section shape survives even at
the end of the propagation distance when the pulse intensity decreases more than ten times and the
nonlinear contribution to the spin wave dynamics should considerably diminish.

In order to interpret the experimental result we have assumed that the development of nonlinear
instabilities in a laterally confined medium is strongly modified by a quantization of the spin wave
spectrum. That is why we have transformed the two-dimensional Nonlinear Schr\"{o}dinger Equation
traditionally used for the analysis of bullet dynamics \cite{Collision_properties_quasi-one-dimensional_spin_wave_solitons_two-dimensional_spin_wave_bullets}
into a system of coupled equations for
amplitudes of the spin wave width modes. The specific form of the discrete set of these orthogonal
modes is defined by the actual boundary conditions at the lateral edges of the stripe. We developed
a two-dimensional theory of linear spin-wave dynamics in magnetic stripes. As an important
outcome we found that the Guslienko-Slavin's effective boundary condition \cite{Guslienko} for dynamic
magnetization at the stripe lateral edges, being initially derived for spin waves with vanishing longitudinal
wavenumbers, is also valid in the case of propagating width modes with non-vanishing
longitudinal wavenumbers 
\cite{APL2007kostylev}
. The effective boundary condition shows that the magnetization vector
at the lateral stripe edges is highly pinned, that means that the amplitude of dynamic magnetization
practically vanishes at the edges. For simplicity it is even possible to consider the stripe width
modes to be totally pinned at the stripe lateral edges. As seen from Fig.~\ref{PulseProfiles} this conclusion is in a
good agreement with the experiment.

The analysis of the system of nonlinear equations derived from the Nonlinear Schr\"{o}dinger Equation
shows that the formation of the two-dimensional waveform can be considered as an enrichment
of the spectrum of the width modes. The partial waveforms carried by the modes have the same
carrier frequencies equal to that of the initial signal and the carrier wave numbers which satisfy
the dispersion relations for the modes. In the linear regime all the modes are orthogonal to each
other and do not interact. In the nonlinear (high amplitude) regime the width modes become intercoupled
by the four-wave nonlinear interaction, resulting in an intermodal energy transfer and the
mode spectrum enrichment.

As the spin wave input antenna effectively generates only the lowest width mode, the initial waveform
launched in the stripe is determined by it solely. Therefore to understand the underlaying
physics of quasi-bullet formation it is necessary to consider the nonlinear interaction of higherorder
width modes with it.

Our theoretical analysis shows that the interaction of the lowest width mode ($n=1$) with higherorder
modes is different for odd and even higher order modes. While interacting with even modes,
the lowest width mode plays the role of the pumping wave. This parametrically transfers its energy
to the higher width modes. The interaction is purely parametric and therefore a threshold process.
It needs an initial signal to start the process. This signal usually is a thermally excited mode.
Therefore the amplified waveform needs a large distance of propagation and a group velocity equal
to the velocity of the lowest width mode in order to reach the soliton amplitude level. If there is
a damping of the pumped wave, even modes will never reach an amplitude comparable with that
of the lowest mode. As a result they can contribute to the nonlinear waveform profile only, if the
amplitude of the initial waveform is far beyond the threshold of soliton formation.

Interaction of modes of the same type of symmetry are described by a parametric term as well as
by an additional pseudo-linear (tri-linear) excitation term, playing the role of an external source of
excitation. Such a pseudo-linear excitation is a threshold-free process. In contrast to parametric
processes it does not need an initial amplitude value to start the the process. The pseudo-linear excitation
is possible only due to the effective dipolar pinning of the magnetization at the stripe edges.
If the edge spins were unpinned, the interaction of all the width modes would be purely parametric.

The purely parametric mechanism of developing a transversal instability is typical for the process
of bullet formation from a plane-wave waveform in an unconfined medium, which distinguishes it
from the process of soliton and bullet formation in the waveguide structures.

In contrast, the transverse instability of a wave packet in a confined medium starts as a pseudolinear
excitation of higher-order width modes. This mechanism ensures a rapid growth of the
symmetric n = 3 mode up to the level where the parametric mechanism starts to work. After that
the main mode together with the n = 3 mode are capable to rapidly generate a large set of yet
higher modes through both pseudo-linear and parametric mechanisms.

Our theory shows that the efficiency of both nonlinear interaction mechanisms (parametric and
tri-linear) strongly depends on the group velocity difference of modes and the initial length of the
nonlinear pulse. In larger stripes the group velocities of modes are closer to each other. As a result
the nonlinearly generated higher-order modes longer remain within the pump pulse. If the pulse is
long enough, they reach significant amplitudes and a bullet-like waveform is formed. In narrower
stripes the group velocity difference is larger, and consequently the nonlinearly generated highorder
waveforms leave faster the pumping area. As a result, for the same pulse length, they do not
reach significant amplitudes. The nonlinear steepening results in the transformation of the lowest
mode into a soliton.

\begin{figure}
\begin{center}
	\includegraphics[width=13cm]{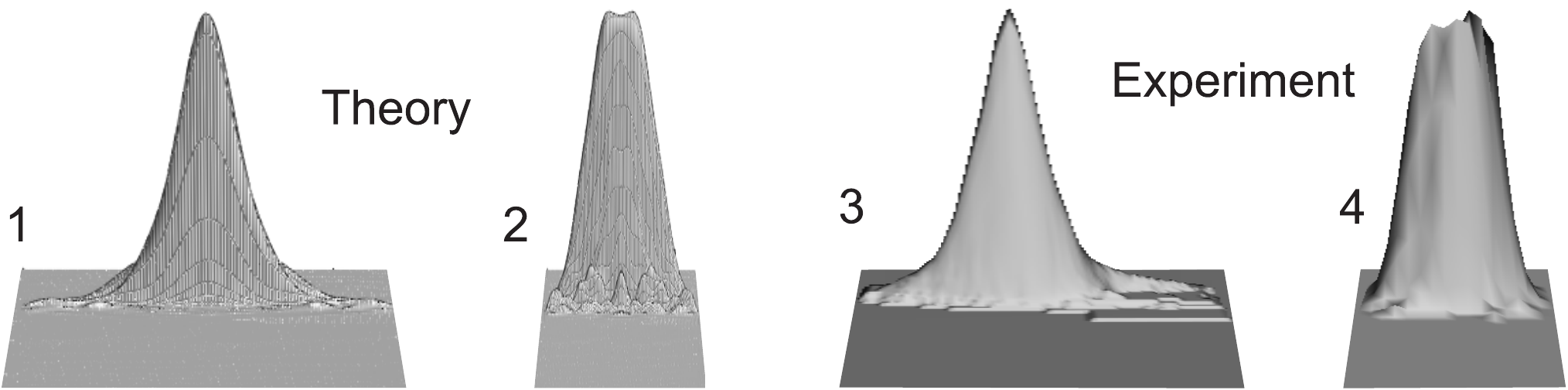} 
	
	 	\caption{Lateral shapes of the nonlinear SW
packets. 1 and 2 -- theoretical results calculated for the ferrite stripes of width of 
2.5mm and 1mm
, respectively. 3 and 4 -- experimental profiles observed in YIG waveguides of width of
2.5mm and 1mm, respectively. 1 and 3: bullets. 2 and 4: solitons.}
		\label{PulseProfiles1}
		\end{center}
\end{figure}

The results of our calculations of the lateral shapes of the nonlinear spin wave packets in wide
(2.5mm)
 and narrow 
1mm 
ferrite stripes are shown in Fig.~\ref{PulseProfiles1}. 
The excellent correspondence with
the experimental data provides good evidence for the validity of the developed theory.

\vspace{.5cm}
Support by the Deutsche Forschungsgemeinschaft, the Australian Research Council, and Russian Foundation for Basic Research is gratefully acknowledged.

\bibliography{}

\end{document}